\documentclass{Interspeech}

\interspeechcameraready 
\usepackage[utf8]{inputenc}
\usepackage{booktabs} 
\usepackage{multirow} 
\usepackage{amsmath}  
\usepackage{graphicx} 
\usepackage{geometry} 
\usepackage{cleveref}
\usepackage{comment}

\title{MeanFlow-TSE: One-Step Generative Target Speaker Extraction with Mean Flow}

\author[affiliation={1}]{Riki}{Shimizu}
\author[affiliation={1}]{Xilin}{Jiang}
\author[affiliation={1}]{Nima}{Mesgarani}

\affiliation{Electrical Engineering}{Columbia University}{United States}
\email{rs4613@columbia.edu}
\keywords{target speaker extraction, speech enhancement, meanflow, one-step generation}

\usepackage{comment}

\begin{document}

\maketitle

\begin{abstract}
    Target speaker extraction (TSE) aims to isolate a desired speaker’s voice from a multi-speaker mixture using auxiliary information such as a reference utterance. Although recent advances in diffusion and flow-matching models have improved TSE performance, these methods typically require multi-step sampling, which limits their practicality in low-latency settings. In this work, we propose MeanFlow-TSE, a one-step generative TSE framework trained with mean-flow objectives, enabling fast and high-quality generation without iterative refinement. Building on the AD-FlowTSE paradigm, our method defines a flow between the background and target source that is governed by the mixing ratio (MR). Experiments on the Libri2Mix corpus show that our approach outperforms existing diffusion- and flow-matching-based TSE models in separation quality and perceptual metrics while requiring only a single inference step. These results demonstrate that mean-flow-guided one-step generation offers an effective and efficient alternative for real-time target speaker extraction. Code is available at \href{https://github.com/rikishimizu/MeanFlow-TSE}{https://github.com/rikishimizu/MeanFlow-TSE}.
\end{abstract}

\section{Introduction}

Target speaker extraction (TSE) isolates the voice of a desired speaker from mixtures of overlapping speech and background noise, guided by speaker-specific cues such as enrollment utterances~\cite{wang_voicefilter_2019, ge_spex_2020}, lip movements~\cite{ephrat_looking_2018, pan_muse_2021}, or direction of arrival~\cite{alcala_padilla_location-aware_2025, he_3s-tse_2024}. By effectively filtering out interfering sounds, TSE is crucial for ensuring robustness in downstream applications such as automatic speech recognition, hearing aids, and telecommunication systems, particularly in acoustically challenging real-world conditions.

Traditional approaches to TSE generally operate as discriminative models that estimate the target speech waveform directly from the input mixture. Typically, models are trained to predict a multiplicative mask applied to learned audio representations or spectrograms to filter out interfering sounds while preserving the target speaker. By conditioning source separation backbones such as Conv-TasNet~\cite{luo_conv-tasnet_2019} or SepFormer~\cite{subakan_attention_2021} with speaker embeddings derived from auxiliary cues, these systems effectively isolate the desired speaker. This discriminative training focuses on minimizing reconstruction errors such as Scale-Invariant Signal-to-Noise Ratio (SI-SNR) to recover the target speaker of interest. While achieving strong signal-level performance, these methods can introduce artifacts and may struggle with generalization to unseen acoustic conditions.

\begin{figure}[t]
  \centering
  \includegraphics[width=\linewidth]{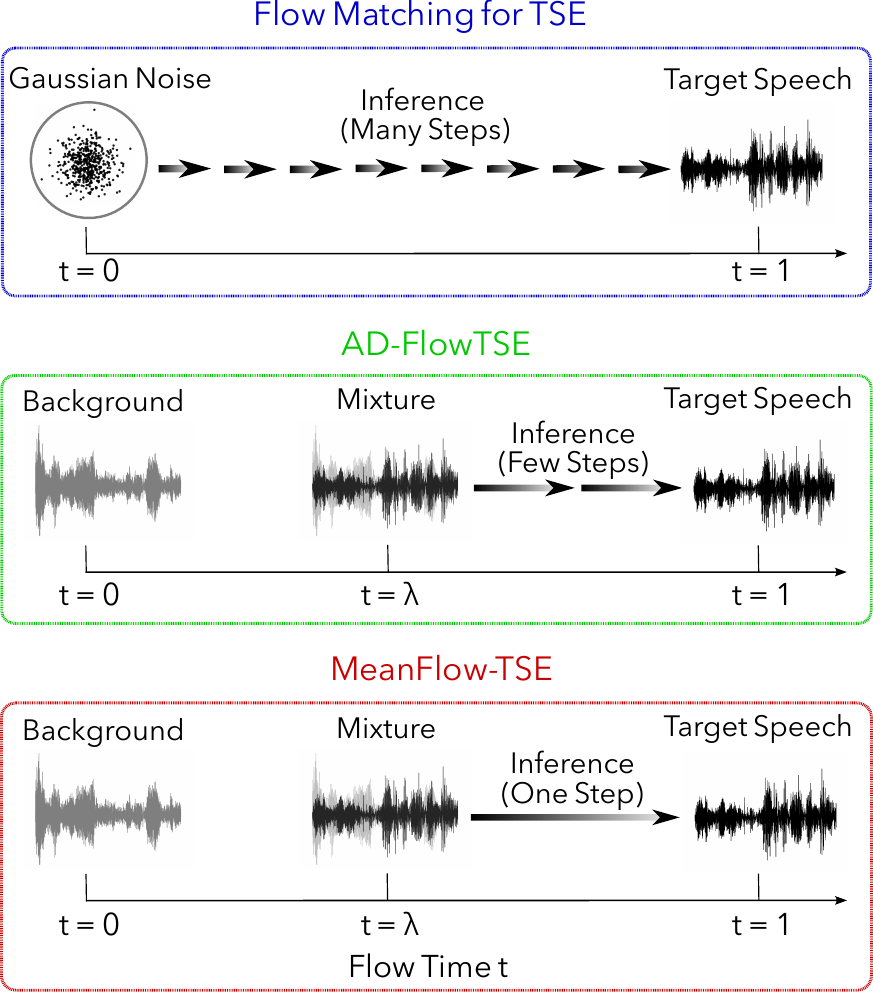}
  \caption{Comparison of generative TSE inference paradigms. \textbf{Top:} Standard Flow Matching transforms the Gaussian noise to the target speech using iterative ODE integration (many steps). \textbf{Middle:} AD-FlowTSE \cite{hsieh_adaptive_2025} defines the flow trajectory between the background and target, initializing inference at the mixture's mixing ratio ($t=\lambda$) to reduce sampling steps. \textbf{Bottom:} The proposed MeanFlow-TSE leverages mean-flow objectives to enable direct, high-quality \textit{one-step} generation from the mixture ($t=\lambda$) to the target speech ($t=1$).}
  \label{fig:fig1}
\end{figure}

Recent work on speech enhancement, separation, and extraction with diffusion models~\cite{ho_denoising_2020} and flow matching~\cite{lipman_flow_2023} represents a paradigm shift from discriminative mask estimation to generative modeling, treating TSE as a conditional generation task. Diffusion-based TSE models~\cite{kamo_target_2023, zhang_ddtse_2024} demonstrated the potential of score-based generative modeling, achieving high perceptual quality. However, these methods suffer from high computational costs and slow inference speeds due to the iterative reverse sampling process, typically requiring at least 10 function evaluations (NFEs) for high-quality outputs. Reducing NFEs is critical for meeting latency constraints in real-time applications such as hearing aids and live telecommunication systems. Flow matching offers a promising alternative by learning straighter probability paths. FlowTSE~\cite{navon_flowtse_2025} applied conditional flow matching to mel-spectrogram domain TSE, while AD-FlowTSE~\cite{hsieh_adaptive_2025} introduced a mixing-ratio-aware transport between background and target sources, enabling adaptive few-step generation. Notably, AD-FlowTSE achieves competitive performance even with a single sampling step, demonstrating the potential of one-step generation for TSE. However, it uses the standard rectified flow-matching objective, originally designed for multi-step sampling, and thus leaves room for improvement for its one-step inference capability.

Mean-flow objectives offer a principled solution for direct one-step generation. Diverging from standard flow matching, which models instantaneous velocities at discrete points along a trajectory, MeanFlow~\cite{geng_mean_2025} models the average velocity across the entire trajectory. This formulation obviates the need for small-step ODE integration, thereby enabling effective one-step generation that achieves state-of-the-art performance on image generation~\cite{geng_mean_2025}, speech enhancement~\cite{zhu_meanflowse_2025, wang_meanse_2025}, and text-to-speech synthesis~\cite{wang_intmeanflow_2025}. This approach was further refined by the $\alpha$-Flow method~\cite{zhang_alphaflow_2025}, which decomposes the mean-flow objective into trajectory flow matching and trajectory consistency terms. By identifying a strong negative correlation between these terms, the authors proposed a curriculum learning strategy that gradually interpolates the objective from trajectory flow matching to the mean-flow identity. Notably, $\alpha$-Flow eliminates the need for Jacobian-vector product calculations during training, significantly reducing computational overhead while enhancing training stability.
Despite these successes in other domains, mean-flow objectives have not yet been explored for target speaker extraction.

In this work, we propose MeanFlow-TSE, the first application of mean-flow objectives to target speaker extraction. Building on the AD-FlowTSE paradigm, our method defines a flow between background and target source governed by the mixing ratio used to create the mixture, enabling mixing-ratio-aware initialization and training. By training with mean-flow objectives, we achieve fast and high-quality one-step generation. Experiments on Libri2Mix demonstrate that MeanFlow-TSE outperforms existing diffusion- and flow-matching-based TSE models in separation quality (SI-SDR) and perceptual metrics (PESQ, ESTOI) while maintaining computational efficiency suitable for real-time applications, establishing mean-flow-guided one-step generation as an effective alternative for practical target speaker extraction systems.

\section{Background}

\subsection{Target Speaker Extraction}
The goal of Target Speaker Extraction (TSE) is to recover the discrete speech waveform $\mathbf{s} \in \mathbb{R}^L$ of a specific target speaker from a single-channel mixture $\mathbf{y} \in \mathbb{R}^L$. The mixture is modeled as the sum of the target speech and an interference component $\mathbf{b} \in \mathbb{R}^L$ (comprising other speakers and background noise):
\begin{equation}
    \mathbf{y} = \mathbf{s} + \mathbf{b}
\end{equation}
TSE systems utilize an auxiliary reference cue $\mathbf{e}$, typically a pre-recorded enrollment utterance, to identify the target speaker. The problem is formulated as estimating $\hat{\mathbf{s}} = f_\theta(\mathbf{y}, \mathbf{e})$. While traditional methods often estimate masks, the generative approach models the trajectory in the spectral domain. Let $\text{STFT}(\cdot)$ denote the Short-Time Fourier Transform. We define the spectral representations of the mixture, target, and background as $\mathbf{Y} = \text{STFT}(\mathbf{y})$, $\mathbf{S} = \text{STFT}(\mathbf{s})$, and $\mathbf{B} = \text{STFT}(\mathbf{b})$, respectively.

\subsection{Flow Matching for TSE}
Flow matching~\cite{lipman_flow_2023} provides a simulation-free framework for training Continuous Normalizing Flows (CNFs). It models the generation process as a deterministic Ordinary Differential Equation (ODE) governing a state $\mathbf{z}_t$ in the latent space (here, the complex spectral domain):
\begin{equation}
    d\mathbf{z}_t = v_\theta(\mathbf{z}_t, t, \mathbf{e}) dt, \quad t \in [0, 1]
\end{equation}
where $v_\theta$ is a neural network predicting the velocity field. The objective is to transform a source distribution $p_0(\mathbf{z})$ to the data distribution $p_1(\mathbf{z})$.
Optimal Transport (OT) defines a straight-line probability path:
\begin{equation}
    \mathbf{z}_t = (1 - t)\mathbf{z}_0 + t\mathbf{z}_1
\end{equation}
The standard flow matching objective minimizes the regression error between the network output and the path derivative:
\begin{equation}
    \mathcal{L}_{\text{FM}}(\theta) = \mathbb{E}_{t, \mathbf{z}_0, \mathbf{z}_1, \mathbf{e}} \left[ \left\Vert v_\theta(\mathbf{z}_t, t, \mathbf{e}) - (\mathbf{z}_1 - \mathbf{z}_0) \right\Vert^2 \right]
\end{equation}

\subsection{Adaptive Deterministic Flow Matching}
AD-FlowTSE~\cite{hsieh_adaptive_2025} adapts this framework by defining the flow between the background representation $\mathbf{B}$ and the target representation $\mathbf{S}$. Although the physical mixture is additive ($\mathbf{y} = \mathbf{s} + \mathbf{b}$), assuming scale invariance of the extraction task, it models the mixture as a convex combination governed by a scalar mixing ratio $\lambda \in [0, 1]$ to align it with the optimal transport trajectory:
\begin{equation}
    \mathbf{y} = \lambda \mathbf{s} + (1 - \lambda) \mathbf{b}
\end{equation}
In the spectral flow domain, the path is defined such that $\mathbf{z}_0 = \mathbf{B}$ and $\mathbf{z}_1 = \mathbf{S}$:
\begin{equation}
    \mathbf{z}_t =  t \mathbf{S} + (1 - t) \mathbf{B}
\end{equation}
where $t \in [0, 1]$. This allows for \textit{mixing-ratio-aware initialization}. Since the mixture spectrogram $\mathbf{Y}$ lies at $t=\lambda$ on this trajectory, inference can be initialized at $\mathbf{z}_{\hat{\lambda}} \approx \mathbf{Y}$ using an estimated mixing ratio $\hat{\lambda}$, reducing the integration range to $[\hat{\lambda}, 1]$.

\subsection{MeanFlow and $\alpha$-Flow Objectives}
Extracting the target speaker effectively via the standard flow matching or diffusion requires multiple function evaluations (NFEs) and ODE steps. To mitigate this computational bottleneck, MeanFlow~\cite{geng_mean_2025} enables one-step generation by modeling the average velocity required to jump between time steps $t$ and $r$ (where $r > t$). The state $\mathbf{z}_r$ is recovered via:
\begin{equation}
    \mathbf{z}_r = \mathbf{z}_t + (r - t) v_{\text{avg}}(\mathbf{z}_t, t, r)
\end{equation}
To improve training efficiency, the $\alpha$-Flow framework~\cite{zhang_alphaflow_2025} defines a hybrid target velocity $v_{t,r}^\alpha$ that interpolates between the ground-truth trajectory and a self-consistency target: Let the ground truth direction be $\mathbf{u} = \mathbf{S} - \mathbf{B}$. The target velocity $v_{t,r}^\alpha$ is defined as:
\begin{equation}
    v_{t,r}^\alpha = \alpha \mathbf{u} + (1 - \alpha) v_\theta(\mathbf{z}_\tau, \tau, r)
\end{equation}
where $\alpha \in (0, 1]$ is a scalar hyperparameter and $\tau = \alpha r + (1 - \alpha)t$ is an intermediate time. The objective minimizes the distance to this interpolated target:
\begin{equation}
    \mathcal{L}_{\alpha}(\theta) = \mathbb{E}_{t, r, \mathbf{B}, \mathbf{S}} \left[ \left\Vert v_\theta(\mathbf{z}_t, t, r) - v_{t,r}^\alpha \right\Vert^2 \right]
\end{equation}

\section{MeanFlow-TSE}

\subsection{Proposed Formulation and Inference}
We adapt MeanFlow to TSE by conditioning the velocity estimator on the enrollment $\mathbf{e}$. The network $v_\theta(\mathbf{z}_t, t, r, \mathbf{e})$ predicts the transport from time $t$ to target time $r$. 

At inference, we aim to recover the clean target $\mathbf{S}$ (where $r=1$) in a single step. Using an estimated mixing ratio $\hat{\lambda}$, we initialize the flow at the mixture $\mathbf{z}_{start} = \mathbf{Y}$ (corresponding to $t_{start} = \hat{\lambda}$). The target spectrogram $\hat{\mathbf{S}}$ is estimated as:
\begin{equation}
    \hat{\mathbf{S}} = \mathbf{Y} + (1 - \hat{\lambda}) v_\theta(\mathbf{Y}, \hat{\lambda}, 1, \mathbf{e})
\end{equation}
This effectively skips the noise-dominant trajectory, jumping directly from the mixture to the target.

\subsection{Adaptive Weight Loss}
Following the $\alpha$-flow framework \cite{zhang_alphaflow_2025}, we employ an adaptive weight $w$ parametrized by $\alpha$ and $c$. Let $\Delta = v_\theta(\mathbf{z}_t, t, r, \mathbf{e}) - v_{t,r}^\alpha$ be the error vector. The weight is defined as:
\begin{equation}
    w = \frac{\alpha}{||\Delta||_2^2 + c}
\end{equation}
where $c=10^{-3}$. The training objective applies this weight with a stop-gradient operator $\text{sg}(\cdot)$:
\begin{equation}
    \mathcal{L}_{\text{Adaptive}}(\theta) = \mathbb{E}_{t, r, \dots} \left[ \text{sg}(w) \cdot ||\Delta||_2^2 \right]
\end{equation}

\subsection{Alpha Curriculum Scheduling}
We employ a curriculum learning strategy that gradually transitions the training objective from trajectory flow matching ($\alpha = 1$) to the mean-flow identity ($\alpha \rightarrow 0$). The hyperparameter $\alpha$ governs both the target velocity interpolation in Eq. (8) and the adaptive weight in Eq. (11). We implement a sigmoid-based schedule over training iterations $k$:
\begin{equation}
    \alpha(k) = \text{clip}\left(1 - \sigma\left(\gamma \cdot \left(\frac{k - k_s}{k_e - k_s} - 0.5\right)\right), \alpha_{\text{min}}, 1.0 \right)
\end{equation}
where $k_s$ and $k_e$ denote the start and end of the transition, $\gamma$ controls the steepness, and $\sigma(\cdot)$ is the sigmoid function. This creates three training phases: (1) trajectory flow matching pretraining ($k < k_s$, $\alpha = 1$), (2) curriculum transition ($k_s \leq k < k_e$, $\alpha$ decreases from 1 to $\alpha_{\text{min}}$), and (3) mean-flow fine-tuning ($k \geq k_e$, $\alpha \approx \alpha_{\text{min}}$). We set $k_s = 0$, $k_e$ at epoch 2000, $\gamma = 25.0$, and $\alpha_{\text{min}} = 0.005$.

\subsection{Mixing Ratio Predictor}
Since $\lambda$ is unknown during inference, we use an auxiliary network $g_\phi$ to estimate it. Using ECAPA-TDNN~\cite{desplanques_ecapa-tdnn_2020} as a feature extractor $w(\cdot)$ and an MLP $h(\cdot)$, the predictor outputs:
\begin{equation}
    \hat{\lambda} = g_\phi(\mathbf{y}, \mathbf{e}) = \sigma\left(h\left([w(\mathbf{y}); w(\mathbf{e})]\right)\right)
\end{equation}
This is trained separately via MSE loss: $\mathcal{L}_{\text{MR}}(\phi) = \mathbb{E} [ (\hat{\lambda} - \lambda)^2 ]$.

\begin{table*}[t]
\centering
\caption{Performance comparison of generative TSE models on the Libri2Mix Noisy and Clean sets.}
\setlength{\tabcolsep}{4pt}
\resizebox{\textwidth}{!}{%
\begin{tabular}{lccccccccccccc}
\toprule
\multirow{2}{*}{Method} & \multirow{2}{*}{Type} & \multicolumn{6}{c}{Libri2Mix Noisy} & \multicolumn{6}{c}{Libri2Mix Clean} \\
\cmidrule(lr){3-8} \cmidrule(lr){9-14}
 & & PESQ & ESTOI & SI-SDR & OVRL & DNSMOS & SIM & PESQ & ESTOI & SI-SDR & OVRL & DNSMOS & SIM \\
\midrule
Mixture & -- & 1.08 & 0.40 & -1.93 & 1.63 & 2.71 & 0.46 & 1.15 & 0.54 & 0.00 & 2.65 & 3.41 & 0.54 \\
\midrule
DiffSep+SV \cite{zhang_ddtse_2024} & \multirow{5}{*}{G} & 1.32 & 0.60 & -- & 2.78 & 3.63 & 0.62 & 1.85 & 0.79 & -- & 3.14 & \textbf{3.83} & 0.83 \\
DDTSE \cite{zhang_ddtse_2024} & & 1.60 & 0.71 & -- & 3.28 & 3.74 & 0.71 & 1.79 & 0.78 & -- & \textbf{3.30} & 3.79 & 0.73 \\
DiffTSE \cite{kamo_target_2023} & & -- & -- & -- & -- & -- & -- & 3.08 & 0.80 & 11.28 & -- & -- & -- \\
FlowTSE \cite{navon_flowtse_2025} & & 1.86 & 0.75 & -- & \textbf{3.30} & \textbf{3.82} & 0.83 & 2.58 & 0.84 & -- & 3.27 & 3.79 & 0.90 \\
SR-SSL \cite{ku_generative_2025} & & -- & -- & -- & -- & -- & -- & 2.99 & -- & 16.00 & -- & -- & -- \\
SoloSpeech \cite{wang_solospeech_2025} & & 1.89 & 0.78 & 11.12 & -- & 3.76 & -- & -- & -- & -- & -- & -- & -- \\
AD-FlowTSE \cite{hsieh_adaptive_2025} & & 2.15 & 0.81 & 12.69 & 3.11 & 3.48 & \textbf{0.87} & 2.89 & 0.90 & 17.49 & 3.15 & 3.59 & \textbf{0.95} \\ 
\midrule
\textbf{MeanFlow-TSE} & \multirow{1}{*}{G} & \textbf{2.21} & \textbf{0.82} & \textbf{12.85} & 3.17 & 3.55 & 0.73 & \textbf{3.26} & \textbf{0.93} & \textbf{18.80} &  3.21 & 3.69 & 0.92\\
\midrule
\end{tabular}%
}
\label{tab:performance}
\end{table*}

\section{Experiments}

\subsection{LibriMix Dataset}

Similar to AD-FlowTSE, we adopted the setup described in \cite{zmolikova_speakerbeam_2019, delcroix_improving_2020} using the Libri2Mix dataset. We combined the \texttt{train-360} and \texttt{train-100} subsets for training, while reserving the \texttt{dev} and \texttt{test} sets for validation and evaluation, respectively. All audio was sampled at 16 kHz. Our input pipeline utilized 6-second segments, composed of a 3-second enrollment speech paired with a 3-second mixture. For spectral feature extraction, we applied an STFT with a hop size of 128 and a window length of 510.

\subsection{Training Configuration}

\subsubsection{Model and Optimization}
The backbone architecture is a U-Net style Diffusion Transformer (UDiT) \cite{liu_generative_2024} comprising 16 transformer layers \cite{vaswani_attention_2017} with 16 attention heads and a hidden dimension of 768. Both input and output dimensions are set to 512, with a positional length of 500 and no positional encodings applied. We optimized the model using AdamW \cite{loshchilov_decoupled_2019} with a weight decay of 0.01. The learning rate followed a cosine annealing schedule ($T_{\text{max}}=50$) with a 5-epoch warm-up, decaying from an initial $1 \times 10^{-4}$ to a minimum of $1 \times 10^{-5}$.

\subsubsection{Training Setup}
The model was trained for 2,000 epochs with a batch size of 32 using distributed data parallelism across eight NVIDIA L40 GPUs. We employed 16-bit mixed-precision training to reduce memory consumption and improve computational efficiency. To prevent exploding gradients, gradient clipping was applied with a threshold of 0.5. During inference, the Euler solver was executed with a single step for both validation and testing, except for NFE analysis in \Cref{fig:fig2}. The final model was selected based on the checkpoint that achieved the highest SI-SDR on the validation set.

\subsubsection{MeanFlow Configuration}

The time steps $r$ and $t$ were sampled from a log-normal distribution parametrized by $\mu=-0.4$ and $\sigma=1.0$. We utilize a flow ratio of $0.5$: essentially, for $50\%$ of the training iterations, the condition $t=r$ is enforced, effectively reducing the objective to standard rectified flow matching. To balance the loss components, the adaptive weight coefficient $c$ was set to $1 \times 10^{-3}$.

\subsection{Baseline Models}

We evaluate MeanFlow-TSE against a diverse set of generative TSE baselines, including DiffSep+SV~\cite{zhang_ddtse_2024}, DDTSE~\cite{zhang_ddtse_2024}, DiffTSE~\cite{kamo_target_2023}, FlowTSE~\cite{navon_flowtse_2025}, SR-SSL~\cite{ku_generative_2025}, and SoloSpeech~\cite{wang_solospeech_2025}. For these models, we adopt the comprehensive benchmark results reported by Navon et al.~\cite{navon_flowtse_2025} under both clean and noisy configurations. Additionally, we include AD-FlowTSE~\cite{hsieh_adaptive_2025} as a primary point of comparison. As AD-FlowTSE shares the same backbone architecture and hyperparameters as our method, it serves as a direct control to isolate the specific efficacy of the MeanFlow paradigm.

\subsection{Evaluation Metrics}

To comprehensively assess model performance, we employ 6 metrics covering speech quality, intelligibility, and speaker similarity. Signal fidelity and perceptual quality are measured using Scale-Invariant Signal-to-Distortion Ratio (SI-SDR)~\cite{roux_sdr_2019}, Perceptual Evaluation of Speech Quality (PESQ)~\cite{rix_perceptual_2001}, and Extended Short-Time Objective Intelligibility (ESTOI)~\cite{taal_short-time_2010}. We also report DNSMOS~\cite{reddy_dnsmos_2021} and OVRL scores to estimate mean opinion scores. Finally, to evaluate how well the target identity is maintained, we calculate Speaker Similarity (SIM)~\cite{wang_wespeaker_2023} based on the cosine similarity between the target and estimated embeddings.

\begin{figure}[t]
  \centering
  \includegraphics[width=\linewidth]{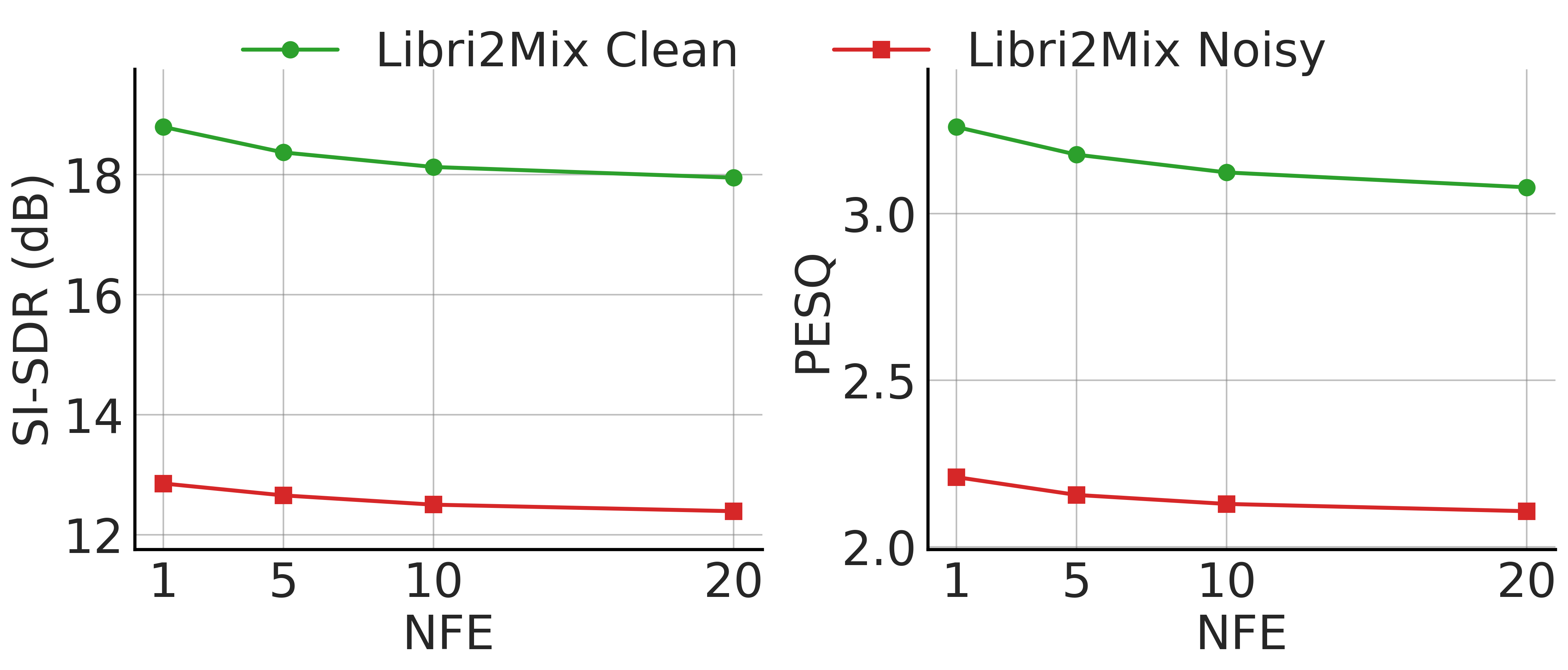}
  \caption{Comparison of performance metrics (SI-SDR and PESQ) across different numbers of function evaluations.}
  \label{fig:fig2}
\end{figure}

\section{Results}

\subsection{Overall Performance}

In \Cref{tab:performance}, we report the performance of MeanFlow-TSE compared against other generative baseline TSE models on the noisy and clean subsets of Libri2Mix.

Overall, our approach consistently outperforms competing generative baselines in terms of extraction and reference-based perceptual quality. On the \texttt{Libri2Mix Clean} test set, MeanFlow-TSE achieves a PESQ of 3.26 and an SI-SDR of 18.80 dB, surpassing the strong AD-FlowTSE baseline by 0.37 and 1.31 dB, respectively. This substantial improvement highlights the efficacy of the mean-flow objective in learning a more accurate one-step transport map compared to the standard rectified flow matching used in AD-FlowTSE. Similar trends are observed on the \texttt{Libri2Mix Noisy} set, where our method achieves state-of-the-art results for intrusive metrics with 2.21 PESQ and 12.85 dB SI-SDR.

Regarding non-intrusive perceptual metrics like OVRL and DNSMOS, which assess speech quality without a ground-truth reference, MeanFlow-TSE achieves competitive results but trails slightly behind FlowTSE and DDTSE. This aligns with the trade-off between intrusive metrics and non-intrusive "naturalness" scores previously reported in \cite{hsieh_adaptive_2025}. Crucially, MeanFlow-TSE improves upon AD-FlowTSE across both OVRL and DNSMOS, effectively narrowing the gap with multi-step generative models while maintaining significantly higher extraction quality.

\begin{table}[t]
  \centering
  \caption{Comparison of Real Time Factor (RTF) and Peak GPU Memory (VRAM) on Libri2Mix. NFE values for baselines correspond to the optimal configurations reported in their respective original studies \cite{wang_solospeech_2025, hsieh_adaptive_2025}.}
  \label{tab:efficiency}
  \resizebox{\columnwidth}{!}{
  \begin{tabular}{lcccc}
    \toprule
    Method & NFE & Params (M) & RTF & Mem (MB) \\
    \midrule
    SoloSpeech \cite{wang_solospeech_2025} & 50 & 589 & 0.75 & 3738 \\
    \midrule
    \multirow{2}{*}{AD-FlowTSE \cite{hsieh_adaptive_2025}} & 1 & \multirow{2}{*}{358} & 0.017 & \multirow{2}{*}{1531} \\
     & 5 & & 0.047 & \\
    \midrule
    \textbf{MeanFlow-TSE} & 1 & 359 & 0.018 & 1536 \\
    \bottomrule
  \end{tabular}
  }
\end{table}

\subsection{NFE Analysis}
To verify the effectiveness of our one-step generation strategy, we analyze the impact of the Number of Function Evaluations (NFE) on extraction performance. \Cref{fig:fig2} illustrates that MeanFlow-TSE achieves its peak performance at NFE = 1 both in terms of SI-SDR and PESQ. This behavior empirically validates the theoretical motivation behind MeanFlow-TSE: the training objective successfully rectifies the flow trajectory, enabling the model to map the noisy mixture directly to the clean target in a single step. Consequently, additional Euler integration steps are not only unnecessary but slightly detrimental, likely due to accumulated discretization errors.

\subsection{Computational Cost Analysis}

\Cref{tab:efficiency} benchmarks the computational efficiency of MeanFlow-TSE against its most competitive baselines, SoloSpeech and AD-FlowTSE. All measurements were conducted on a single NVIDIA L40 GPU using a batch size of 1 and 3-second audio segments. 

As a diffusion-based model, SoloSpeech requires 50 Number of Function Evaluations (NFE), resulting in a significantly higher Real-Time Factor (RTF) and memory footprint than its flow-based counterparts. In contrast, MeanFlow-TSE and AD-FlowTSE utilize a near-identical backbone architecture; while MeanFlow-TSE includes a small additional time-step embedder, the resulting increase in parameter count and peak memory usage is marginal. Crucially, these results demonstrate that the performance gains reported in \Cref{tab:performance} are achieved with negligible computational overhead relative to AD-FlowTSE. With its single-step inference and minimal footprint, MeanFlow-TSE is uniquely suited for real-time deployment on edge devices.

\section{Conclusions}

We introduced MeanFlow-TSE, a novel framework for one-step generative target speaker extraction. Our method achieves state-of-the-art SI-SDR, PESQ, and ESTOI performance on the Libri2Mix benchmark. Future work will extend this framework to reverberant environments and multi-channel scenarios.

\bibliographystyle{IEEEtran}
\bibliography{MeanFlow}

\end{document}